\documentclass[
onecolumn
]{aastex62}
\usepackage{graphics} 
\usepackage{epsfig} 
\usepackage{mathptmx} 
\usepackage{times} 
\usepackage{amsmath} 
\usepackage{amssymb}  
\usepackage{xcolor}
\usepackage{subfigure}
\usepackage{bm}
\usepackage{hyperref}
\usepackage{chngcntr}

\graphicspath{{./}}

\submitjournal{ApJ}

\shorttitle{Correlation of CME Shock Temperature with SEP Intensity}
\shortauthors{Cuesta et al.}

\begin{document}

\title{\bf Correlation of Coronal Mass Ejection Shock Temperature with Solar Energetic Particle Intensity}

\correspondingauthor{Manuel Enrique Cuesta}
\email{mecuesta@princeton.edu}

\author[0000-0002-7341-2992]{Manuel Enrique Cuesta}
\affiliation{Princeton University,
Department of Astrophysical Sciences,
Princeton, NJ 08544, USA}

\author[0000-0001-6160-1158]{D. J. McComas}
\affiliation{Princeton University,
Department of Astrophysical Sciences,
Princeton, NJ 08544, USA}

\author[0000-0003-0412-1064]{L. Y. Khoo}
\affiliation{Princeton University,
Department of Astrophysical Sciences,
Princeton, NJ 08544, USA}

\author[0000-0002-6962-0959]{R. Bandyopadhyay}
\affiliation{Princeton University,
Department of Astrophysical Sciences,
Princeton, NJ 08544, USA}

\author[0000-0002-8527-1509]{T. Sharma}
\affiliation{Princeton University,
Department of Astrophysical Sciences,
Princeton, NJ 08544, USA}

\author[0000-0002-3093-458X]{M. M. Shen}
\affiliation{Princeton University,
Department of Astrophysical Sciences,
Princeton, NJ 08544, USA}

\author[0000-0002-8111-1444]{J. S. Rankin}
\affiliation{Princeton University,
Department of Astrophysical Sciences,
Princeton, NJ 08544, USA}

\author{A. T. Cummings}
\affiliation{Princeton University,
Department of Astrophysical Sciences,
Princeton, NJ 08544, USA}

\author[0000-0003-2685-9801]{J. R. Szalay}
\affiliation{Princeton University,
Department of Astrophysical Sciences,
Princeton, NJ 08544, USA}

\author[0000-0002-0978-8127]{C. M. S. Cohen}
\affiliation{Space Research Lab, California Institute of Technology, Pasadena, CA 91125, USA}

\author[0000-0002-3737-9283]{N. A. Schwadron}
\affiliation{Princeton University,
Department of Astrophysical Sciences,
Princeton, NJ 08544, USA}
\affiliation{University of New Hampshire, Durham, NH 03824, USA}

\author[0000-0002-7174-6948]{R. Chhiber}
\affiliation{NASA Goddard Space Flight Center, Greenbelt, MD 20771, USA}
\affiliation{Department of Physics and Astronomy, Bartol Research Institute, University of Delaware, Newark, DE 19716, USA}

\author[0000-0003-4168-590X]{F. Pecora}
\affiliation{Department of Physics and Astronomy, Bartol Research Institute, University of Delaware,  Newark, DE 19716, USA}

\author[0000-0001-7224-6024]{W. H. Matthaeus}
\affiliation{Department of Physics and Astronomy, Bartol Research Institute, University of Delaware,  Newark, DE 19716, USA}

\author[0000-0002-0156-2414]{R. A. Leske}
\affiliation{Space Research Lab, California Institute of Technology, Pasadena, CA 91125, USA}

\author[0000-0002-7728-0085]{M. L. Stevens}
\affiliation{Smithsonian Astrophysical Observatory, Cambridge, MA 02138, USA}









\begin{abstract}
    Solar energetic particle (SEP) events have been observed by the Parker Solar Probe (PSP) spacecraft since its launch in 2018.
    These events include sources from solar flares and coronal mass ejections (CMEs).
    Onboard PSP is the IS\(\odot\)IS instrument suite measuring ions over energies from \(\sim\)~20~keV/nucleon to 200~MeV/nucleon and electrons from \(\sim\)~20~keV to 6~MeV.
    Previous studies sought to group CME characteristics based on their plasma conditions and arrived at general descriptions with large statistical errors, leaving open questions on how to properly group CMEs based solely on their plasma conditions.
    To help resolve these open questions, plasma properties of CMEs have been examined in relation to SEPs.
    Here we reexamine one plasma property, the solar wind proton temperature, and compare it to the proton SEP intensity in a region immediately downstream of a CME-driven shock for seven CMEs observed at radial distances within 1~au.
    We find a statistically strong correlation between proton SEP intensity and bulk proton temperature, indicating a clear relationship between SEPs and the conditions in the solar wind.
    Furthermore, we propose that an indirect coupling of SEP intensity to the level of turbulence and the amount of energy dissipation that results is mainly responsible for the observed correlation between SEP intensity and proton temperature.
    These results are key to understanding the interaction of SEPs with the bulk solar wind in CME-driven shocks and will improve our ability to model the interplay of shock evolution and particle acceleration.
    \newline
\end{abstract}



\section{Introduction} \label{sec:intro}

Coronal mass ejections (CMEs) are solar events involving the expulsion of large amounts of solar coronal material into interplanetary space.
These events are observed to have wide variability in their characteristics, such as their composition, magnetic field strength and behavior, in-situ plasma conditions, and shock strengths.
Many of these characteristics have been studied resulting in an array of expectations that vary from one CME to the next \citep[][and references therein]{CaneEA1990JGR_CME_SEP_obs,GoslingEA1991JGR_IPshock_CME_1au,NeugebauerEA1997JGR_CME_Tails,ZurbuchenRichardson2006SSRv_ICME_signatures,RichardsonCane2010SoPh_ICME_summaries,WuLepping2011SoPh_ICME_clouds, ChiEA2016SoPh_ICME_stats}.
A recent investigation into the turbulence properties of CME cores inside 1~au \citep{GoodEA2023ApJL_Turb_ICMEs} via Solar Orbiter and Parker Solar Probe (PSP) observations revealed radial trends in the magnetic turbulent energy and the inertial-to-dissipative range break-point frequency, which is analogous to the dissipation scale where turbulent energy cascaded from larger to smaller scales can dissipate as heat.
These authors further demonstrated the variability of other CME characteristics, such as plasma beta (ratio of thermal to magnetic pressure) and correlation length, leaving an open question on how to properly group CMEs based solely on their plasma properties.

To better group CMEs, their plasma properties have been examined in relation to solar energetic particles (SEPs).
CMEs can drive an upstream forward propagating shock if they are traveling faster than the upstream solar wind by at least the local fast magnetosonic speed, where the upstream plasma is unshocked and the downstream plasma is shocked.
These CME-driven shocks become a source for particle acceleration \citep{Lee1997GMS_CMEshock_transport} with variable efficiency in their acceleration processes \citep{Kallenrode1996JGR_Shock_intensity_efficiency,LarioEA1998ApJ_efficiency}, where turbulent diffusion is thought to produce the shock's dominant particle acceleration process \citep{Fisk1971JGR_CosRayIntensity_IPshock,AxfordEA1977ICRC_CosRayAccel_Shock,BlandfordOstriker1978ApJL_ParticleAccel_Shock,Bell1978MNRAS_ShockAccel_CosRay,GoslingEA1979AIPC_IonAccel_EarthBowShock,Lee1983JGR_IonAccel_IPshock}.
Noticeable increases in the suprathermal particle population (energies generally between \(\sim\)~10~keV to \(\sim\)~1~MeV per nucleon \citep{MasonGloeckler2012SSRv_suprathermals}) and of those with higher energy at CME-driven shocks are the result of these local acceleration processes \citep{Reames1999SSRv_particle_accel}. 
CME-driven shock speeds have been found to be correlated with peak SEP flux at the shock \citep{CaneEA1990JGR_CME_SEP_obs,Reames2000AIPC_CME_intensity_speed,Kahler2001JGR_CME_intensity_speed} for higher energies (\(>10\)~MeV), with considerable scatter.
This correlation can be explained by the association of stronger shocks with more intense CMEs resulting in a larger SEP flux of higher energy particles.
Nonetheless, the considerable scatter over several decades of SEP intensity for a given range of CME speeds suggests the peak SEP flux is also dependent on other factors besides the CME speed, such as the ambient SEP population \citep{Kahler2001JGR_CME_intensity_speed}.
When considering the behavior of the peak energetic particle (EP) intensity in the heliosheath, \citet{ZankEA2015ApJ_DSA_reconnection_Heliosheath} found that the EP intensity peaks downstream of the heliospheric bow shock and the distance between the shock and this peak increases with increasing particle energy.

One method used to characterize the relationship between SEPs and turbulence is the partial variance of increments (PVI) \citep{GrecoEA2008}, a measure of turbulence intermittency or roughness of the magnetic field at a given increment scale, which is found to be well-correlated with solar wind temperature \citep{OsmanEA2011ApJL_PVI_Temp,Osman2012PhRvL_PVI_Temp,GrecoEA2012,ChasapisEA2015ApJL_PVI_Temp,YordanovaEA2016GeoRL_Temp_structures,QudsiEA2020ApJS_PVI_Temperature}.
Additionally, \citet{TesseinEA2013ApJL_PVI_intensity,TesseinEA2015ApJ_EPs_PVI} and \citet{BandyopadhyayEA2020ApJS_PVI_intensity} found a correlation between PVI and SEP intensity, where others discovered an association between local acceleration regions and magnetic-like islands formed due to turbulence \citep{KhabarovaEA2015ApJ_MagIslands_acceleration,KhabarovaZank2017ApJ_EPs_CurrentSheets,MalandrakiEA2019ApJ_EPs_CurrentSheets}.
Although PVI and SEP intensity are correlated on average, there are instances where SEP enhancements are observed at times with low PVI.
These times with larger SEP intensity, however, are bounded by high PVI, suggesting that coherent structures play a role in the local modulation and trapping of energetic particles \citep{TesseinEA2015ApJ_EPs_PVI,Tessein2016GRL_EPtrapping_CoherentStructures,PecoraEA2021MNRAS_SEP_helicity}.

Based on the outcomes of the observational studies above, one might expect SEP intensity can be used to gauge the solar wind proton temperature, a connection motivated by the relation between SEP intensity and shock strength.
From the perspective of theory and modeling, the influence of the upstream turbulence level and presence of coherent structures are found to result in more efficient diffusive-like spreading of particles in both position and velocity space \citep{TrottaEA2021PNAS_Model_Shocks_Turbulence}.
Furthermore, the level of upstream turbulence induces shock-surface fluctuations and downstream turbulence structuring that have been linked to temperature anisotropy downstream of the shock \citep{TrottaEA2023MNRAS_3Dmodel_shock_turb}.
However, there are no studies to date that have directly investigated the relationship between the bulk proton temperature immediately downstream of CME-driven shocks and the intensity of suprathermal particle populations locally accelerated by turbulent diffusion near the shock (although, see \citet{DayehEA2018JPhCS_ESP_Property_Variability} and the inferred results of \citet{LarioEA2017JPhCS_EP_Pressures}).
The shock strength and turbulence strength near the shock are found to be well-correlated, where turbulence is known to affect SEP transport \citep{Jokipii1982ApJ_DiffusionAcceleration,Jokipii1987ApJ_DiffusiveShockAccel,Droge1994ApJS_SEP_transport,GuoEA2021FrASS_Turb_SEP_transport}.
This natural interaction between SEPs and turbulence near the shock may also be reflected in the proton temperature, since intermittency results in increased heating of the bulk material at the targeted scales.

In this paper, we examine the relationship of proton temperature and particle intensity through the analysis of SEP and in-situ plasma measurements by PSP \citep{Fox2016SSRv_PSP} for seven CME-driven shocks.
In Section \ref{sec:data_method}, we explain the data and methods used to reach the results in Section \ref{sec:results}. 
We discuss the results and their impact on the understanding of SEP interactions with the bulk solar wind in shocks in Section \ref{sec:discussion}.
Finally, we summarize the results in Section \ref{sec:conclusion}.

\section{Data and Methods} \label{sec:data_method}
    
We utilize publicly available measurements of solar energetic particles via PSP/IS\(\odot\)IS EPI-Lo \citep{McComasEA2016SSRv_ISOIS}, magnetic field data via PSP/FIELDS \citep{BaleEA2016}, and proton temperatures, bulk velocity, and proton density via PSP/SWEAP/SPAN-I \citep{KasperEA2016} from NASA's Space Physics Data Facility\footnote{Space Physics Data Facility can be accessed at \href{https://cdaweb.gsfc.nasa.gov/}{https://cdaweb.gsfc.nasa.gov/}.}.
We identify possible shock candidates from the FIELDS and SWEAP data and identify the relevant one by comparing its time to the estimated shock arrival time.
Estimated shock arrival times are found by first considering their associated type-II radio bursts \citep{BaleEA1999GRL_Type2BurstSource,KnockEA2001JGR_Type2shockTheory,MannKlassen2005AA_ShockWave_ElectronBeam}, measured by PSP/FIELDS, adjusted to solar origin, denoted by time \(t_0\).
The time of solar origin is determined by subtracting the travel time of light from the Sun to the radial distance of PSP from the time of the observed radio-II burst.
Next, we find the time associated with the propagation of the CME-driven shock (estimated by the WSA-ENLIL model \citep{OdstrcilEA1996GRL_ENLIL1,OdstrcilPizzo1999JGR_ENLIL2,OdstrcilEA1999JGR_ENLIL3,Odstrcil2003AdSpR_ENLIL4,OdstrcilEA2004JGRA_ENLIL5} for each shock separately) starting from a solar origin to the radial distance of PSP, denoted as time \(t_1\).
Then we select the nearest shock candidate to the time \(t_0 + t_1\).
This gives us a total of seven CME-driven shocks, overviews of which can be found in the Appendix (Figure \ref{fig:overview}), with vertical dashed red lines used to mark the corresponding shock arrivals.

To reinforce the connection between SEP intensity and proton temperature via turbulent processes, we compute the PVI of the magnetic field downsampled to 1~second cadence from the native `4\_Sa\_per\_Cyc' data product.
The PVI is defined in \citet{GrecoEA2008} as

\begin{equation} \label{eq:PVI}
    {\rm PVI}(t,\tau) = \frac{\left| \Delta \bm{B}(t,\tau)\right|}{\sqrt{\langle \left|\Delta \bm{B}(t, \tau)\right|^2\rangle}}
\end{equation}
where \(\Delta \bm{B}(t,\tau) = \bm{B}(t+\tau) - \bm{B}(t)\) is the increment of magnetic field vector \(\bm{B}\) at time \(t\) and temporal increment scale \(\tau\), and the brackets \(\langle \cdot \rangle\) represent an arithmetic mean.
We choose an increment scale of 10~seconds, which is shorter than that used in \citet{BandyopadhyayEA2020ApJS_PVI_intensity} due to the availability of higher resolution magnetic field data and is in the inertial range far from the correlation scale but larger than the dissipative scale.

The duration of the arithmetic mean used in the normalization of PVI varies from event to event.
To determine this duration, we first compute the auto-correlation function:

\begin{equation} \label{eq:corr}
    R(\tau) = \frac{\langle \bm{b}(t) \cdot \bm{b}(t+\tau) \rangle_t}{\langle \bm{b}(t) \cdot \bm{b}(t) \rangle_t}
\end{equation}
where \(\bm{b}(t) = \bm{B}(t) - \langle \bm{B}(t) \rangle_t \) is the fluctuating component of the magnetic field and \(\langle \cdot \rangle_t\) is computed via the Germano approach \citep{Germano1992JFM_Filtering}.
The interval duration used in Equation \eqref{eq:corr} is set at 16~hours centered about the estimated shock arrival time, which is an appropriate interval duration at 1~au \citep{IsaacsEA2015}.
The value of \(\tau_c\) such that \(R(\tau=\tau_c)\) drops below \(e^{-1}\) is the ``e-folding'' time or correlation time \citep{MatthaeusEA1999AIPC_CorrelationsTurb}.
We then set the duration of the arithmetic mean used in the normalization of PVI to \(10 \cdot \tau_c\) centered about the estimated shock arrival time.

However, in this study, PSP encounters each of the CME-driven shocks within 1~au where the correlation scale is expected to be shorter, on average, which presumably calls for shorter averaging intervals when computing magnetic correlations.
Therefore, assuming a 16~hour duration will lead to an overestimated correlation scale and a larger averaging duration used for normalizing the PVI.
Although this is important with investigations that directly depend on the correlation scale, an averaging duration for the PVI normalization beyond 10 times the actual correlation scale is found to insignificantly affect PVI enhancements \citep{ServidioEA2011JGRA_PVIaveraging_Reconnection}.
Once the PVI(\(\tau=10~{\rm seconds}\)) is obtained, we compute the fraction of PVI values, denoted as PVI\(_{frac}\), greater than 2.5 only within the duration used for time-integration and averaging of the energy-integrated intensity and proton temperature, respectively.
This threshold is selected since coherent structures are associated with PVI values greater than 2.5 \citep{GrecoEA2008,GrecoEA2012,GrecoEA2018}.
 

To compute the SEP intensity of protons shown below in Section \ref{sec:results}, we examine proton intensities measured by the EPI-Lo instrument.
EPI-Lo is comprised of 8 wedges, each with 10 apertures having different look directions \citep{HillEA2017JGRA_EPILo}.
Here we utilize all apertures that are unaffected by dust puncture events on EPI-Lo (\citet{SzalayEA2020ApJS_PSP_DustImpact}; \citet[][in prep.]{ShenEA2023_dust}) to account for the changing magnetic field direction (or changing pitch angle, the angle between the observed particle's incident direction and the instantaneous magnetic field direction) during CME shock passage.
The data product used in this study is Channel P, which corresponds to protons with triple coincidence measurements \citep{HillEA2017JGRA_EPILo}.

We first integrate the measured flux (\(F\)) over the energy (\(E\)) domain (100~keV to 500~keV), normalized by the summation of the energy bin widths to properly account for energy bin weights.
This results in an energy-integrated flux (\(F_E(t)\)) defined as 

\begin{equation} \label{eq:F_E}
    F_E(t) = \frac{\sum^{500 {\rm keV}}_{100 {\rm keV}} \left[ \Delta E_i \cdot F(t,E_i) \right]}{\sum^{500 {\rm keV}}_{100 {\rm keV}} \left[ \Delta E_i \right]}
\end{equation}
where \(\Delta E_i\) is the energy bin width associated with the measured flux at energy \(E_i\).
We then integrate \(F_E(t)\) over the time domain for a duration equivalent to \(3 \cdot \tau_c\), which is expected to vary from event to event.
We choose this duration to (1) constrain the results to describe the downstream region nearest to the shock and (2) include magnetic structures large enough to capture the acceleration of and/or systematic trapping of energetic particles.
For event-to-event comparisons, we normalize time integration by the duration of the integration.
We also confirmed that the qualitative results and strength of correlation between any pair of parameters (proton temperature, proton SEP intensity, and PVI\(_{frac}\); see Appendix, Table \ref{tab:correlations}) are not sensitive to our exact choice of time window; changing the durations of the window from 0.5 to \(5 \cdot \tau_c\) only insignificantly affects the results.

This results in a time and integrated measured flux in units of \({\rm (cm^2 \cdot sr \cdot s \cdot keV/nuc)^{-1}}\), defined as

\begin{equation} \label{F_E_t}
    F_{E,t} = {\rm Intensity} = \frac{\sum^{t_s+3\tau_c}_{t_s}\left[ \Delta t \cdot F_E(t) \right]}{\sum^{t_s+3\tau_c}_{t_s}\left[ \Delta t \right]}
\end{equation}
where \(t_s\) is the time of the CME-driven shock. 
For the rest of the paper, we interchangeably use the terms `flux' and `intensity' to describe the value \(F_{E,t}\) where it does not cause confusion.
We compare the resulting proton SEP intensity to the proton temperature \(T_p\) measured by SPAN-I from PSP/SWEAP in its native time resolution, averaged over the same duration of \(3 \cdot \tau_c\) downstream of the CME-driven shock.

One challenge is that instrument uncertainties in SPAN-I proton temperatures are unavailable.
Therefore, we assume an uncertainty of each measurement of 20\% \citep{KasperEA2016}, which is a reasonable estimate for the radial distances at which PSP observed the CMEs (\(\sim\) 0.35~au to .80~au) and the SPAN-I field-of-view throughout the averaging duration.
Quality flags in the SPAN-I data product are checked to ensure reasonable confidence in the measured value.
Also, we impose one condition on the data for the SEP event to be considered for analysis: data availability throughout the passing of a CME-driven shock by all three instrument suites (IS\(\odot\)IS/FIELDS/SWEAP). 
Several CME-driven shocks, beyond those included here, observed by PSP/IS\(\odot\)IS were rendered unusable for the present analysis due to data gaps from instrument turn-offs and spacecraft rolls for satellite repositioning during the shock encounter. 

\section{Results}\label{sec:results}

Here we compile the shock intensity and proton temperature for seven different CMEs, with event overviews provided in the Appendix, Figure \ref{fig:overview}.
In Figure \ref{fig:shockepoch}, we provide an example of the analysis for the Feb. 15, 2022, CME event.
The top panel shows energy-integrated proton intensities (\(F_E(t)\)) over energies 100~keV to 500~keV, the middle panel shows bulk proton temperatures, and the bottom panel shows PVI at a 10~second increment.
The vertical dashed red line corresponds to the estimated shock arrival time, for which each curve is adjusted. 
The duration between the estimated shock arrival time and the vertical solid black line corresponds to the time window used to compute \(F_{E,T}\) and average proton temperature (in this case, \(3 \cdot \tau_c\)).
Parameters of average proton temperature (\(\langle T_p \rangle\)), proton SEP intensity, and PVI\(_{frac}\) over a \(3\cdot \tau_c\) duration for the seven CME events are given in the Appendix, Table \ref{tab:params}.

\begin{figure}[hb]
    \centering
    \includegraphics[width=.8\textwidth]{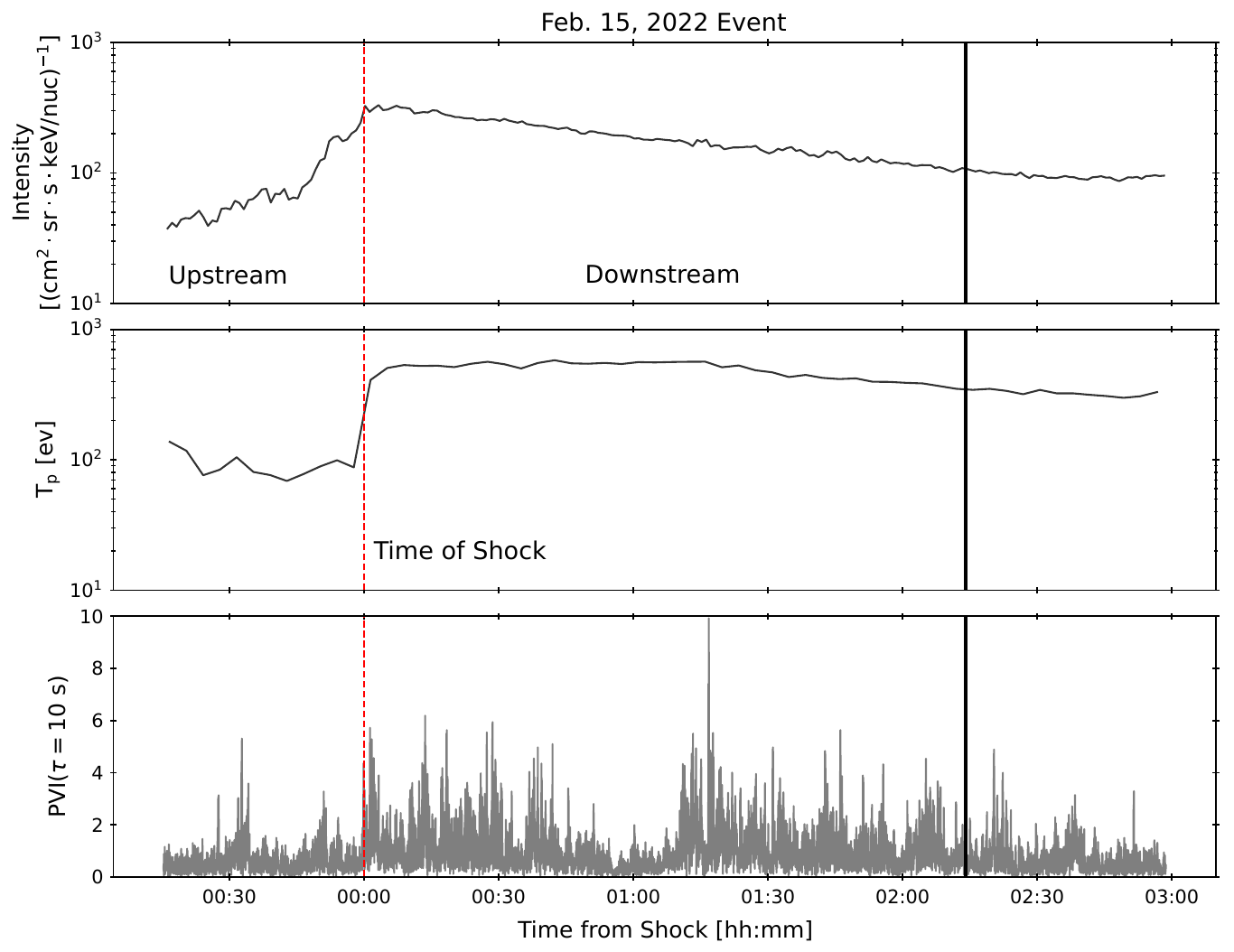}
    \caption{Energy-integrated proton intensity over energies 100~keV to 500~keV (top), solar wind proton temperature (middle), and PVI(\(\tau=10~{\rm seconds}\)) (bottom) centered about a \(3\cdot \tau_c\) duration marked between the shock arrival (vertical dashed red line) and the vertical dashed black line associated with the Feb. 15, 2022, CME event.}
    \label{fig:shockepoch}
\end{figure}

\begin{figure}[ht]
    \centering
    \includegraphics[width=.9\textwidth]{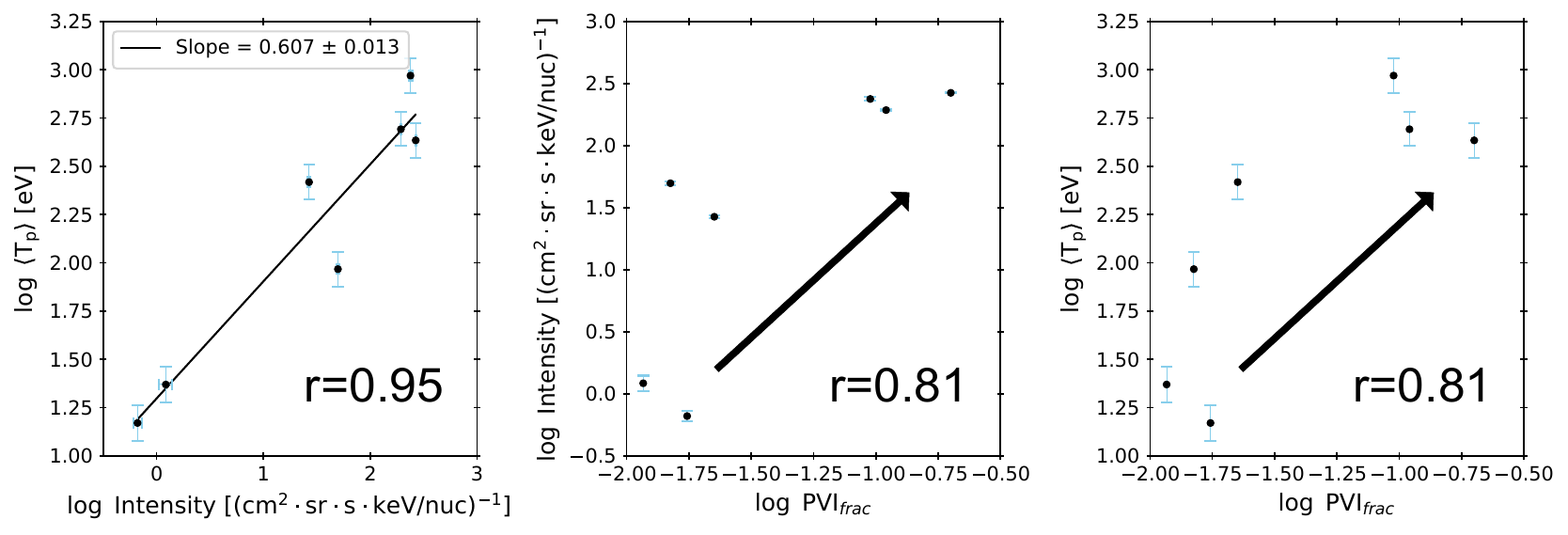}
    \caption{Compiled solar wind \(\langle T_p \rangle\), proton SEP intensity, and PVI\(_{frac}\) each as a function of the other (black circles) in the region immediately downstream of the appropriate CME-driven shock. In the left panel, vertical error bars are propagated standard errors from averaged temperatures and horizontal error bars are propagated uncertainties from time and energy integrated proton flux. The solid black line is an ODR power-law fit to show proportionality. The correlation coefficients (r) of the parameters in the left, middle, and right panel are \(0.95\), \(0.81\), and \(0.81\), respectively. Arrows in the middle and right panels are to express a general trend.}
    \label{fig:temp_v_intensity}
\end{figure}

The compiled values of the intensity and proton temperature are shown in the left panel of Figure \ref{fig:temp_v_intensity} for the selected duration of \(3\times \tau_c\).
The horizontal error bars represent the propagated error throughout the time and energy integrated intensity, and the vertical error bars represent the propagated error from the proton temperatures.
We utilize an orthogonal distance regression \citep[ODR;][]{BoggsRogers1990book_ODRfitting} power-law fit between the proton SEP intensity and \(\langle T_p \rangle\), which considers uncertainty in dependent and independent variables, only to show the proportionality between the two parameters (left panel of Figure \ref{fig:temp_v_intensity}).
The fitted slope and standard error are given in the corresponding legend.
In the middle and right panels of Figure \ref{fig:temp_v_intensity}, we examine the behavior of proton SEP intensity and proton temperature as a function of PVI\(_{frac}\), respectively.
An arrow is drawn in the middle and right panels of Figure 2 only to further demonstrate the relationship between PVI and either \(\langle T_p \rangle\) or SEP intensity; regions with proportionately enhanced PVI are associated with regions of hotter bulk plasma and larger SEP intensity.

Figure \ref{fig:temp_v_intensity} reveals correlations between SEP intensity, bulk proton temperature, and PVI\(_{frac}\) downstream of the shock over a duration of \(3\cdot \tau_c\).
While there is only a small statistical sample via PSP observations due to constraints on data availability, the correlation coefficients (\(r\)) of \(\log \langle T_p \rangle\) to \(\log {\rm Intensity}\), \(\log {\rm Intensity}\) to \(\log {\rm PVI}_{frac}\), and \(\log \langle T_p \rangle\) to \(\log {\rm PVI}_{frac}\) are 0.95, 0.81, and 0.81, respectively, which all correspond to statistically strong correlations \citep{Taylor1990CorrCoeffInterp}.
In the Appendix, Table \ref{tab:correlations}, we provide the correlation coefficients for various durations downstream of the CME-driven shock, which shows insignificant changes in the correlation coefficients between any two parameters.
The results here demonstrate that the proton SEP intensity can be used to gauge the bulk proton temperature in the downstream region of CME-driven shocks.
 
\section{Discussion} \label{sec:discussion}

\begin{figure}[ht]
    \centering
    \includegraphics[width=\textwidth]{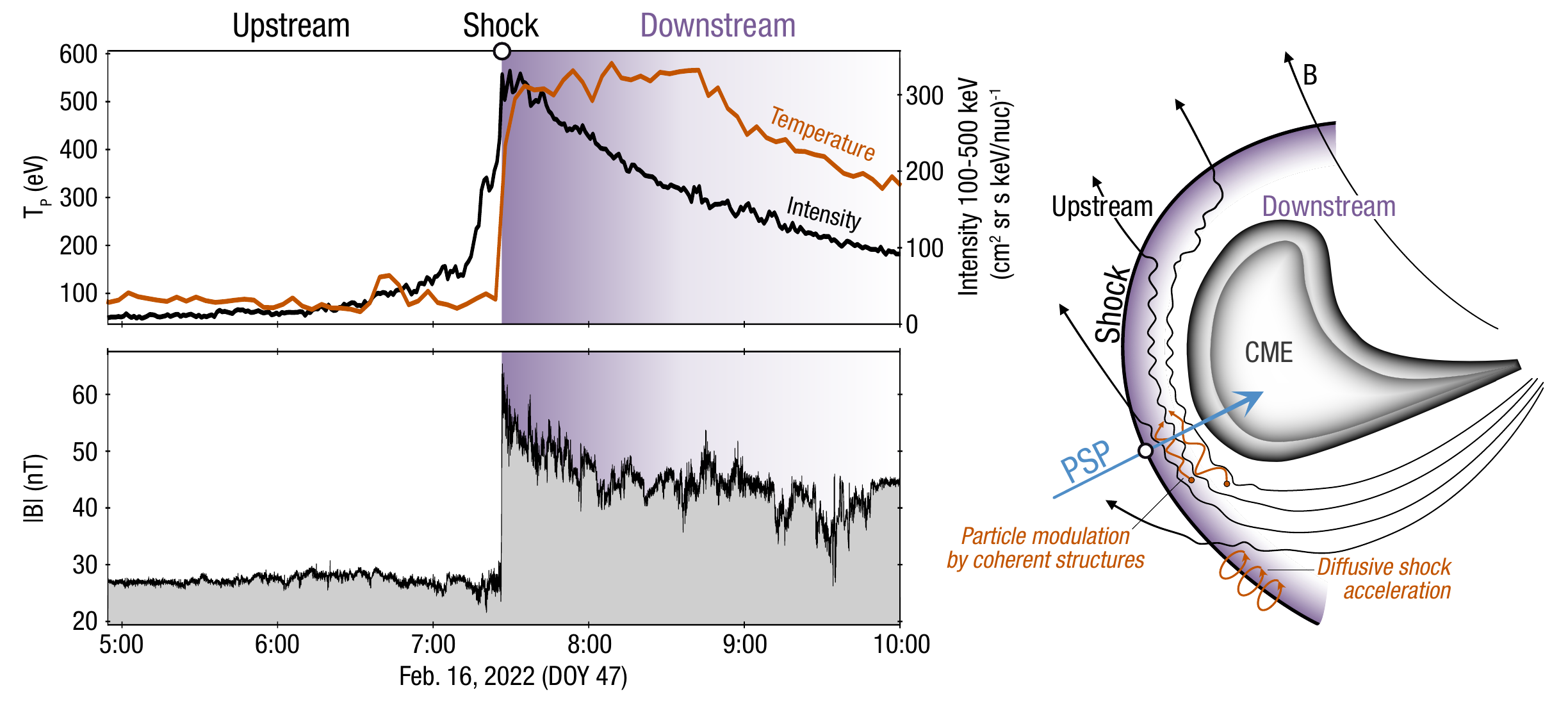}
    \caption{Illustration of CME structure with PSP crossing its driven shock. The top panel shows the measured proton flux integrated over energy (100~keV to 500~keV) and proton temperature; the bottom panel shows the magnetic field magnitude. The CME-driven shock encounter is labeled on PSP's sketched trajectory as an open circle, which corresponds to the open circle on the top axis of the top panel. The trajectory shows PSP traversing the upstream region, encountering the shock, and then observing the downstream plasma. Sketch of CME structure is adapted from \citet{Cane2000SSRv_CME_ForbushDecrease}.}
    \label{fig:illustration}
\end{figure}

We provide an illustration (Figure \ref{fig:illustration}) to help visualize our explanations of the observed results that we discuss below.
Figure \ref{fig:illustration} has real data of a CME-driven shock crossing by PSP for the Feb. 15, 2022, CME event accompanied by an illustration depicting the structure of a CME and a sample trajectory across its associated shock.

An interpretation of the results presented in Figure \ref{fig:temp_v_intensity} involves the level of turbulence and the shock strength. 
Suprathermal particles can be trapped and/or modulated by the emergence of coherent structures, increasing the amount of time they spend in the diffusive shock region, which are visualized in the illustration of Figure \ref{fig:illustration}.
If these particles remain sufficiently trapped, then one might expect a larger intensity of particles observed at that shock.  
This property reflects the acceleration efficiency of the shock and the level of turbulence in these regions, resulting in elevated energy dissipation (heating) due to the emergence of coherent structures, as can be measured via PVI \citep[][and references therein]{QudsiEA2020ApJS_PVI_Temperature}.
In fact, the PVI shown in Figures \ref{fig:shockepoch} and \ref{fig:overview} reveal enhanced values of PVI in the selected region downstream of the shock, corroborating the correlation of PVI and SEP intensity \citep{TesseinEA2013ApJL_PVI_intensity,TesseinEA2015ApJ_EPs_PVI,BandyopadhyayEA2020ApJS_PVI_intensity} and/or the effectiveness of coherent structures in locally modulating/trapping energetic particles \citep{MazurEA2000ApJL_IMFmixing_SEPs,RuffoloEA2003ApJL_SEPtrapping,KittinaradornEA2009ApJL_CoronalLoopHeating,Tessein2016GRL_EPtrapping_CoherentStructures,KhabarovaEA2021SSRv_CurrentSheets_FluxRopes,PezziEA2021SSRv_CurrentSheets_FluxRopes,PecoraEA2021MNRAS_SEP_helicity}.
Additionally, the correlation coefficients between all possible pairs of PVI, proton temperature, and SEP intensity indicate statistically strong correlations. 
As a result, the SEP intensity enables an inference of the proton temperature from the increased heating due to turbulence intermittency.

Supplementing the above interpretation, an increased intensity at the shock can also be explained by the bulk temperature and its impact on the efficiency of particle injection at shocks.
High bulk temperatures increase the likelihood for particles to have sufficient speed to move upstream and thereby become injected into diffusive shock acceleration \citep{GloecklerGeiss1998SSRv_InjectionEff,ChotooEA2000JGR_Temperature_injectionEff,ZurbuchenEA2000AIPC_Temperature_InjectionEff}.
This dynamic can be visualized in Figure \ref{fig:illustration} where particles move towards the upstream plasma and get picked up into diffusive shock acceleration.
This is due to the increased bulk temperature raising the thermal core closer to the suprathermal energies, enabling an increased seed population for acceleration and/or trapping within the diffusion region.
Given the current results, this implies a strong interplay between solar wind temperature, injection efficiency, and SEP intensity.
 
Considerations of other well-correlated properties of CME-driven shocks with SEP intensity, such as shock normal relative to the magnetic field \citep{LarioEA2016ApJ_Intensity_ShockNormal,Giacalone2017ApJ_Intensity_ShockNormal,ChenEA2022ApJ_Intensity_shockNormal}, radial distance from the Sun due to the weakening of such shocks \citep{SalmanEA2020JGRA_Radial_ShockStats}, and connection to the shock \citep{BurlagaEA1981JGR_CMEatSameRadius,KilpuaEA2011JASTP_CME_sameRadius,MostlEA2009SoPh_CME_SameRadius} may influence the present results. 
However, many of these other properties are associated with the shock strength, which is an indicator of the efficiency in acceleration processes due to the turbulent nature downstream of the shock.
Therefore, we only consider the individual connections of the intensity and temperature to the level of turbulence in the downstream region.

\section{Conclusion} \label{sec:conclusion}

In this paper, we investigate the relationship between shock proton temperature with shock associated particle intensity. 
The intensity to temperature correlation had previously only been inferred by existing correlations of PVI, a measure of turbulence intermittency, to both intensity and temperature, separately.
Our results (Figure \ref{fig:temp_v_intensity}) suggest the proton SEP intensity can be used to gauge whether one would expect a larger or smaller proton temperature immediately downstream of a CME-driven shock.
We compute the PVI to help strengthen the explanations relating the SEP intensity to the proton temperature, which reveals enhanced PVI in the downstream region.

The results presented here will likely improve the understanding and modeling of SEP interactions with shocks and the downstream bulk plasma.
Additional spacecraft observations of SEPs and proton temperatures can be used to significantly increase the statistical significance of these results and possibly examine their radial profiles.
The correction for radial weakening of shocks and temperature cooling due to expansion will need to be considered when examining the radial dependence of CME properties in general.
Finally, as the Sun transitions into solar maximum, we anticipate an increase of observed CME-driven shocks in future PSP orbits that can be used to supplement the results presented here in addition supporting other systematic studies of CME-related phenomena.

\section{Acknowledgements}

We thank the IS\(\odot\)IS/FIELDS/SWEAP teams and everyone that made the PSP mission possible. The IS\(\odot\)IS data and visualization tools are available to the community at \href{https://spacephysics.princeton.edu/missions-instruments/PSP}{https://spacephysics.princeton.edu/missions-instruments/PSP}. 
PSP was designed, built, and is operated by the Johns
Hopkins Applied Physics Laboratory as part of NASA’s Living with a Star (LWS) program (contract NNN06AA01C).
R. Chhiber is particularly supported by a NASA LWS grant (80NSSC22K1020).
W. H. Matthaeus and F. Pecora are particularly supported by a PSP/IS\(\odot\)IS subcontract (SUB0000165) and a NASA HGI grant (80NSSC21K1765) to the University of Delaware.
    


\newpage
\appendix

\section{Supplementary Tables and Figures}
\counterwithin{figure}{section}

We consider cases of different durations used in the time-integration and averaging duration of the intensity and proton temperature, respectively.
As shown in Table \ref{tab:correlations}, changing this duration insignificantly alters the strength of correlation between any two parameters.
We also provide the the correlation time and the values of proton temperature, SEP intensity, and PVI\(_{frac}\) from Figure \ref{fig:temp_v_intensity} for each event, as shown in Table \ref{tab:params}.
These values in Table \ref{tab:params} correspond to a time-integration and averaging duration of \(3\cdot \tau_c\) immediately downstream of the CME-driven shock.

\begin{table}[h]
    \centering
    \begin{tabular}{c|c|c|c|c|c|c}
         Correlation  & \(0.5\cdot \tau_c\) & \(1 \cdot \tau_c\) & \(2 \cdot \tau_c\) & \(3\cdot \tau_c\) & \(4\cdot \tau_c\) & \(5\cdot \tau_c\) \\
         \hline
         \hline
         \(\log \langle T_p\rangle\) to \(\log {\rm Intensity}\)              & 0.94 & 0.92 & 0.94 & 0.95 & 0.96 & 0.95 \\
         \(\log {\rm Intensity}\) to \(\log {\rm PVI}_{frac}\) & 0.81 & 0.81 & 0.81 & 0.81 & 0.82 & 0.83 \\
         \(\log \langle T_p\rangle\) to \(\log {\rm PVI}_{frac}\)             & 0.87 & 0.84 & 0.82 & 0.81 & 0.79 & 0.81 \\
    \end{tabular}
    \caption{Correlation coefficients for the pair of parameters given in the first column for different time-integration/averaging duration. Overall, the correlations do not significantly alter across different duration based on factors of the correlation time \(\tau_c\).}
    \label{tab:correlations}
\end{table}

\begin{table}[h]
    \centering
    \begin{tabular}{c|c|c|c|c}
         Event & \(\langle T_p\rangle\) [eV] & \shortstack{SEP Intensity \\ \(\left[({\rm cm^{2} \cdot sr \cdot s \cdot keV/nuc})^{-1}\right]\)} & PVI\(_{frac}\) & \(\tau_c\) [s] \\
         \hline
         \hline
         2020-11-29 & 933 \(\pm\) 194 & 237.8  \(\pm\) 8.9 & 0.09 & 630  \\
         2021-05-28 & 262 \(\pm\) 54  & 26.7   \(\pm\) 0.9 & 0.02 & 3130 \\
         2022-02-15 & 492 \(\pm\) 99  & 193.4  \(\pm\) 2.1 & 0.11 & 2680 \\
         2022-03-21 & 15  \(\pm\) 3   & 0.7    \(\pm\) 0.1 & 0.02 & 8730 \\
         2022-07-09 & 23  \(\pm\) 5   & 1.2    \(\pm\) 0.2 & 0.01 & 2880 \\
         2022-08-28 & 431 \(\pm\) 89  & 266.4  \(\pm\) 3.0 & 0.20 & 1730 \\
         2023-01-03 & 93  \(\pm\) 19  & 49.8   \(\pm\) 1.4 & 0.01 & 1530 \\
    \end{tabular}
    \caption{Values of \(\langle T_p \rangle\), SEP intensity, PVI\(_{frac}\), and \(\tau_c\) for each event using a duration of \(3\cdot \tau_c\) for time-integration of the energy-integrated particle flux and averaging of \(T_p\). These values are reflected in Figure \ref{fig:temp_v_intensity}.}
    \label{tab:params}
\end{table}

Figure \ref{fig:overview} provides an overview of the seven events examined in this study. 
For each event, there are five panels that represent the EPI-Lo Channel P omni-directional proton flux spectrogram, magnetic field, \({\rm PVI}(\tau=10~{\rm seconds})\), magnitude of the bulk velocity, proton density, and proton temperature in descending order. 
The time range includes the estimated shock arrival time, marked by the vertical dashed red line.
The time window used for integration of the energy-integrated intensity and averaging of the proton temperature is between the shock arrival time and the time marked by the vertical solid red line.
These event overviews provide confirmation of data availability throughout these time periods.

\begin{figure}[!htp]
    \centering
    \gridline{
    \includegraphics[width=.5\textwidth]{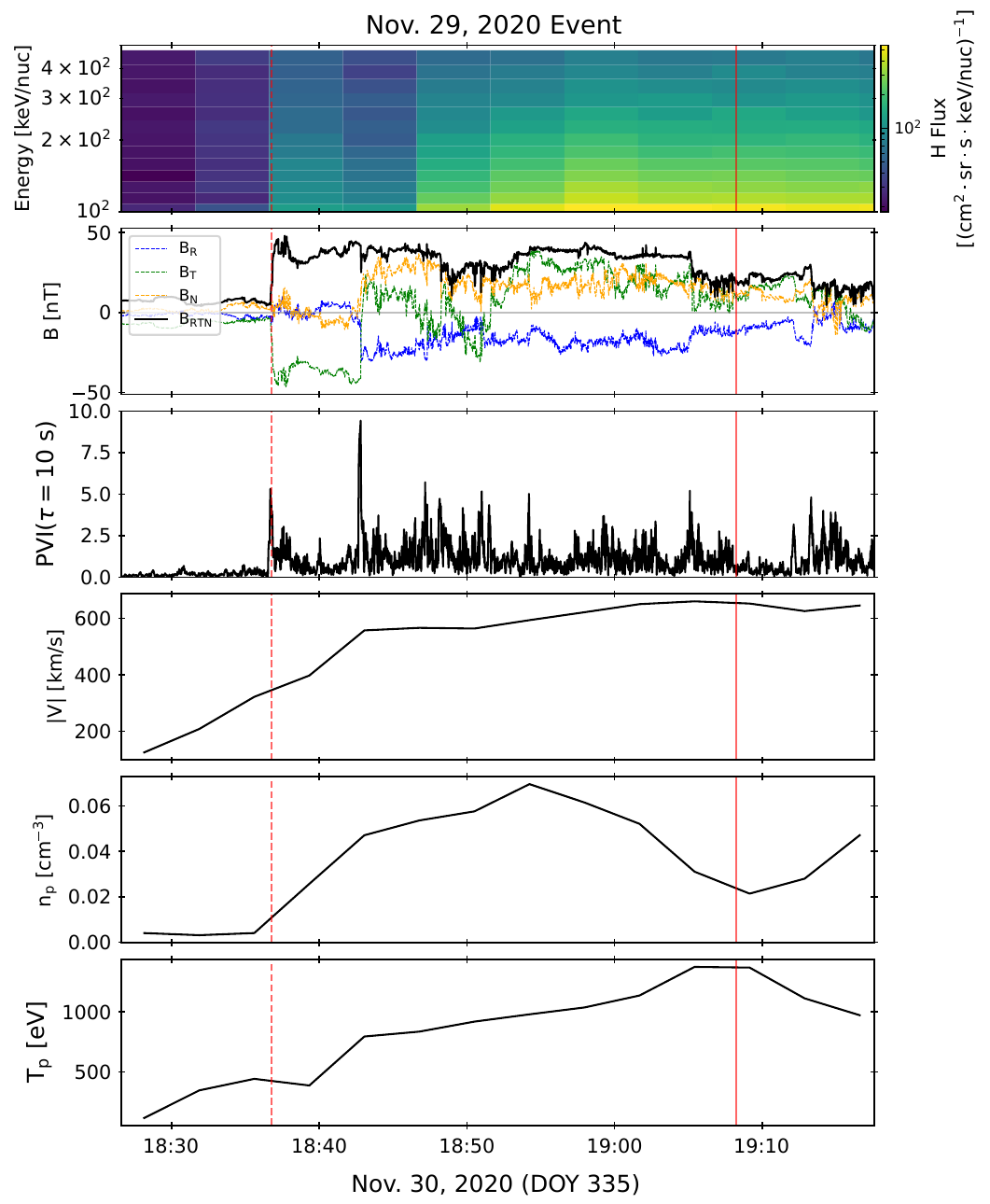}
    \includegraphics[width=.5\textwidth]{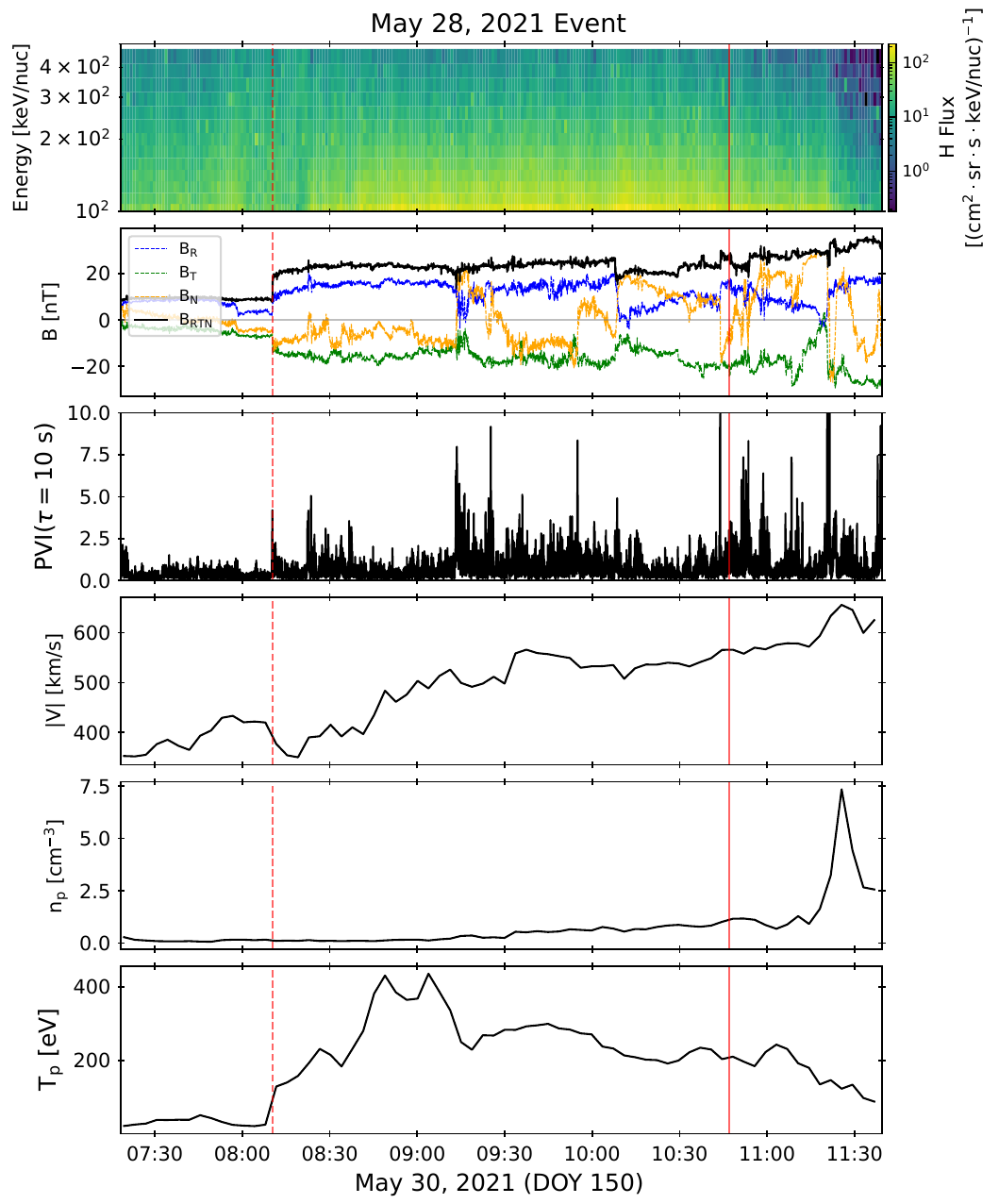}
    }
    \vspace{.1in}
    \gridline{
    \includegraphics[width=.5\textwidth]{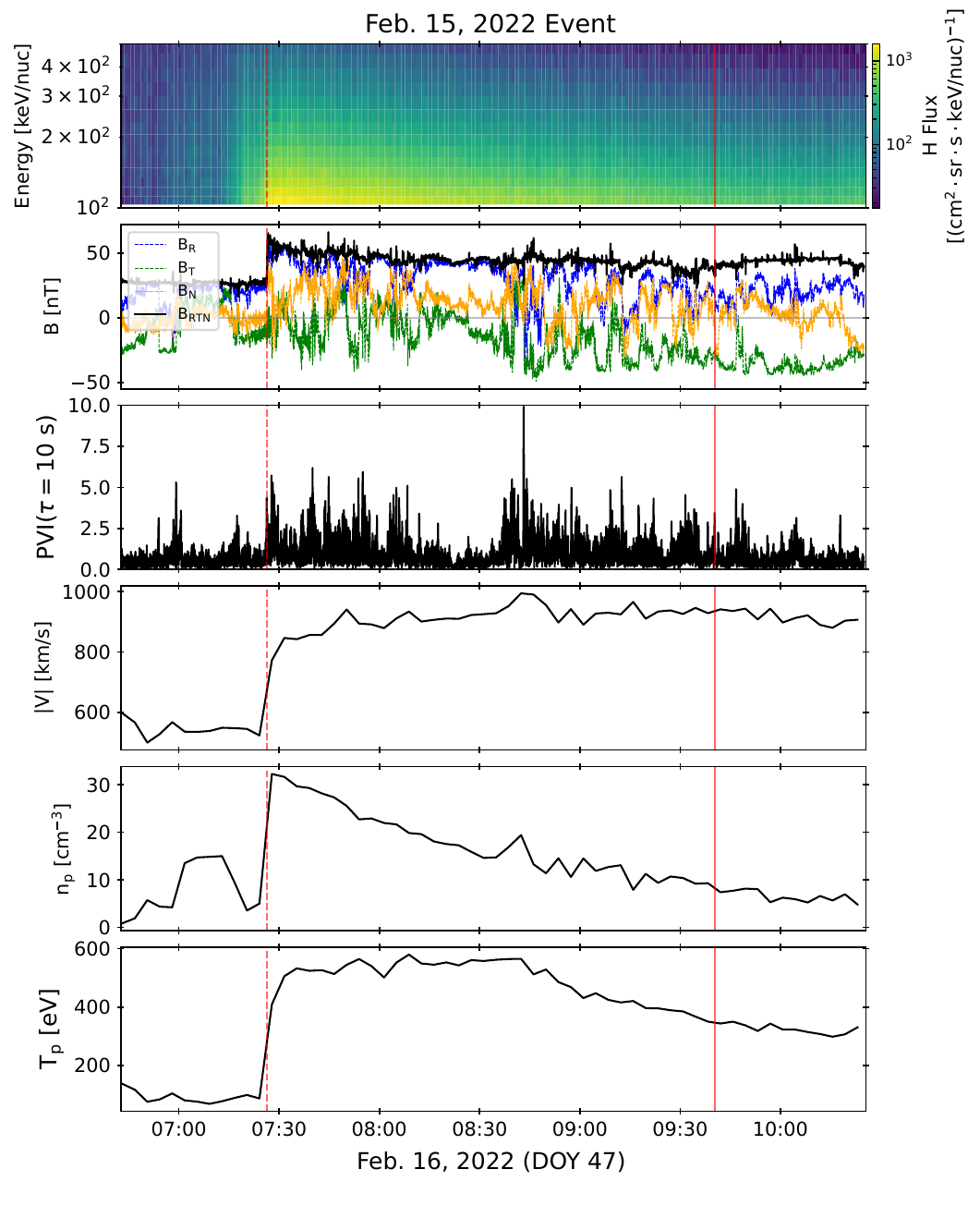}
    \includegraphics[width=.5\textwidth]{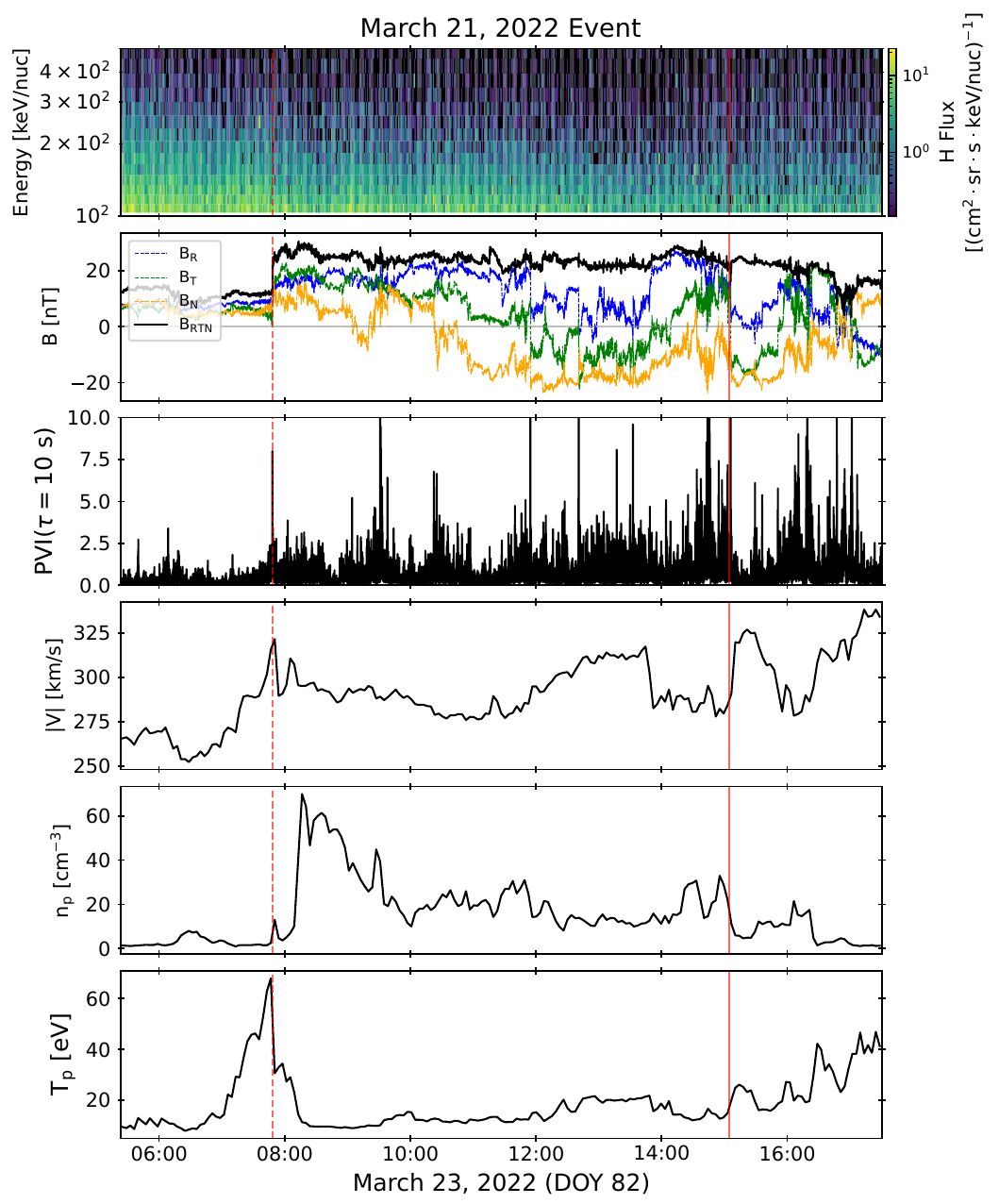}
    }
\end{figure}

\begin{figure}[!htp]
    \centering
    \gridline{
    \includegraphics[width=.5\textwidth]{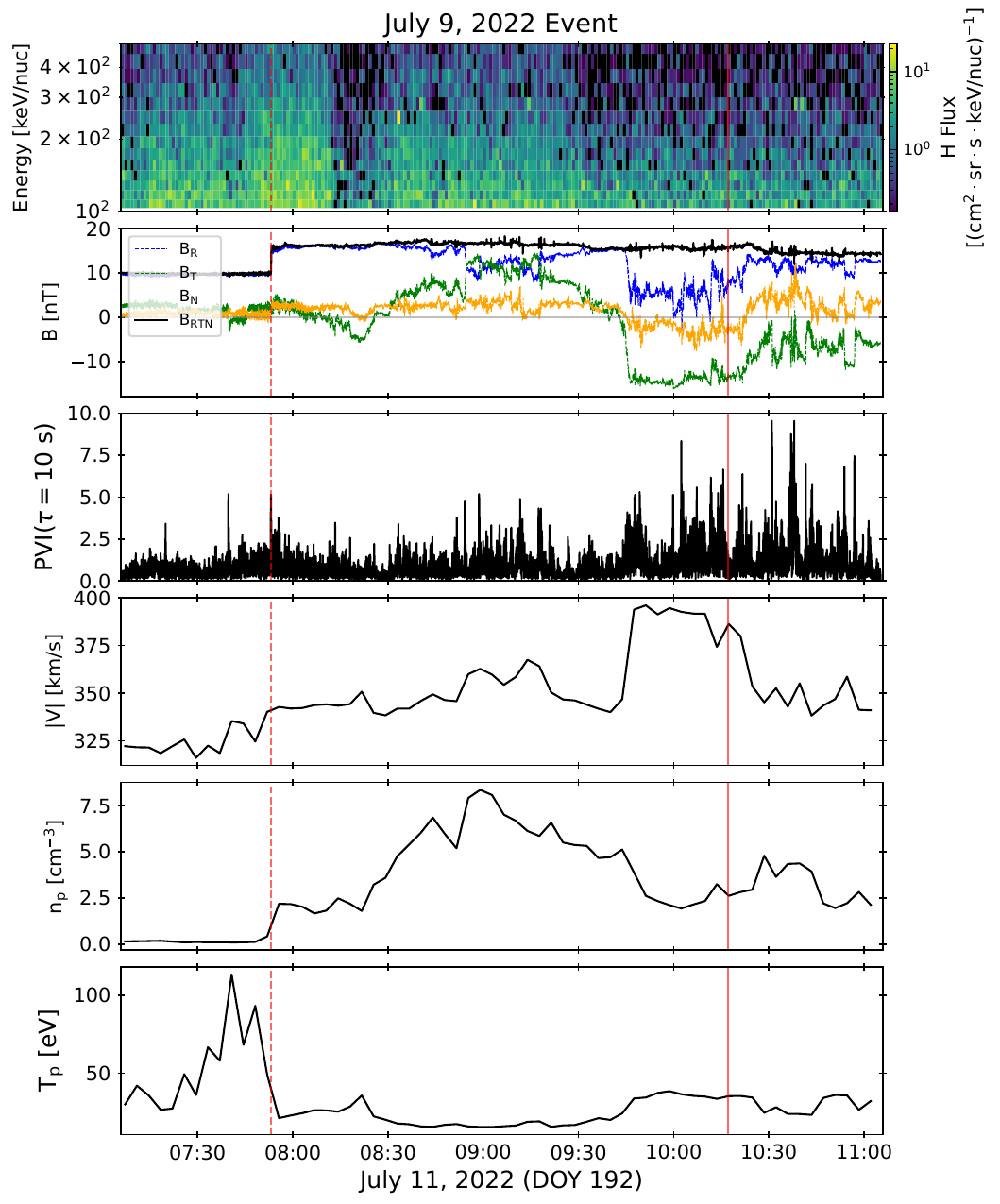}
    \includegraphics[width=.5\textwidth]{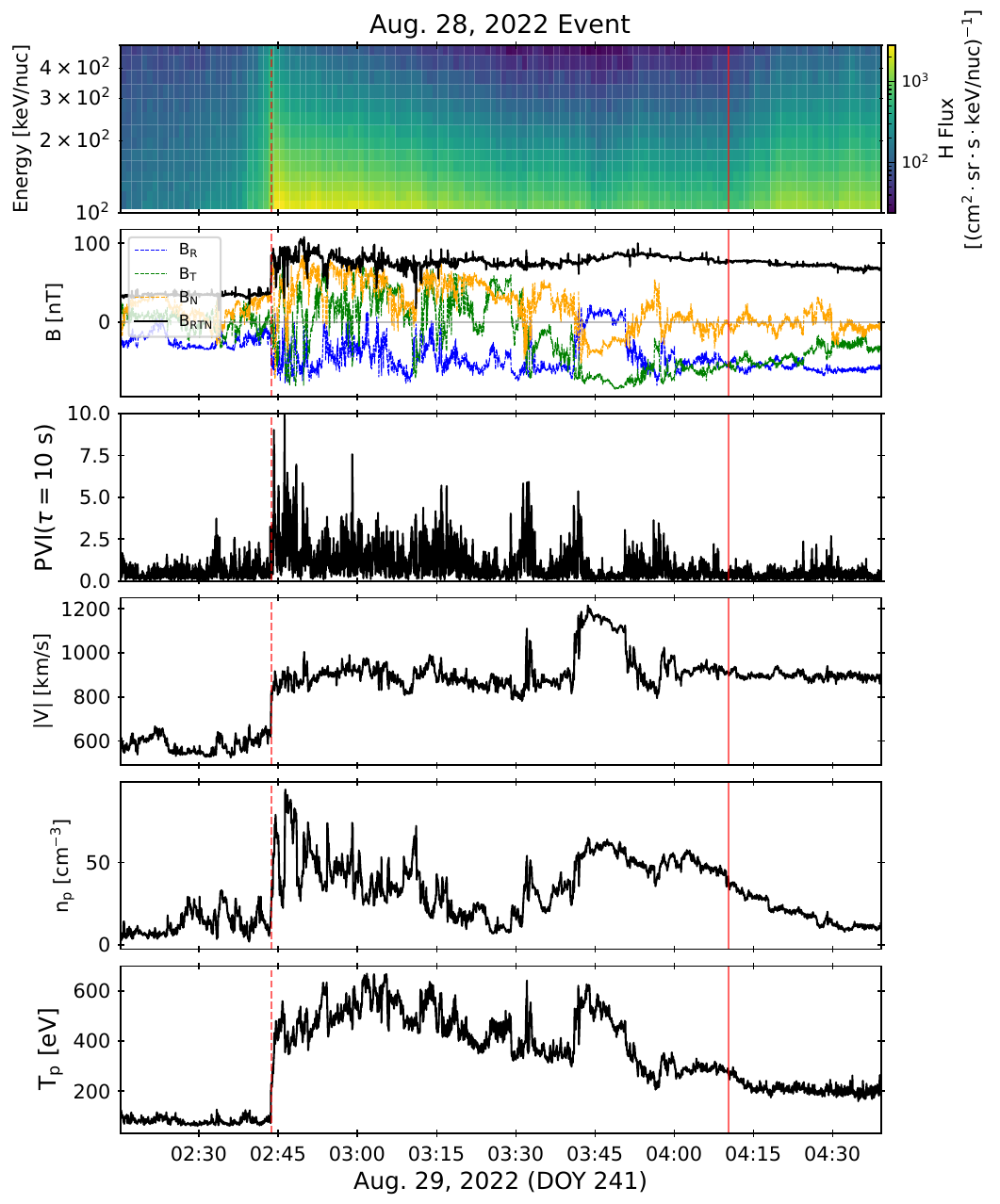}
    }
    \includegraphics[width=.5\textwidth]{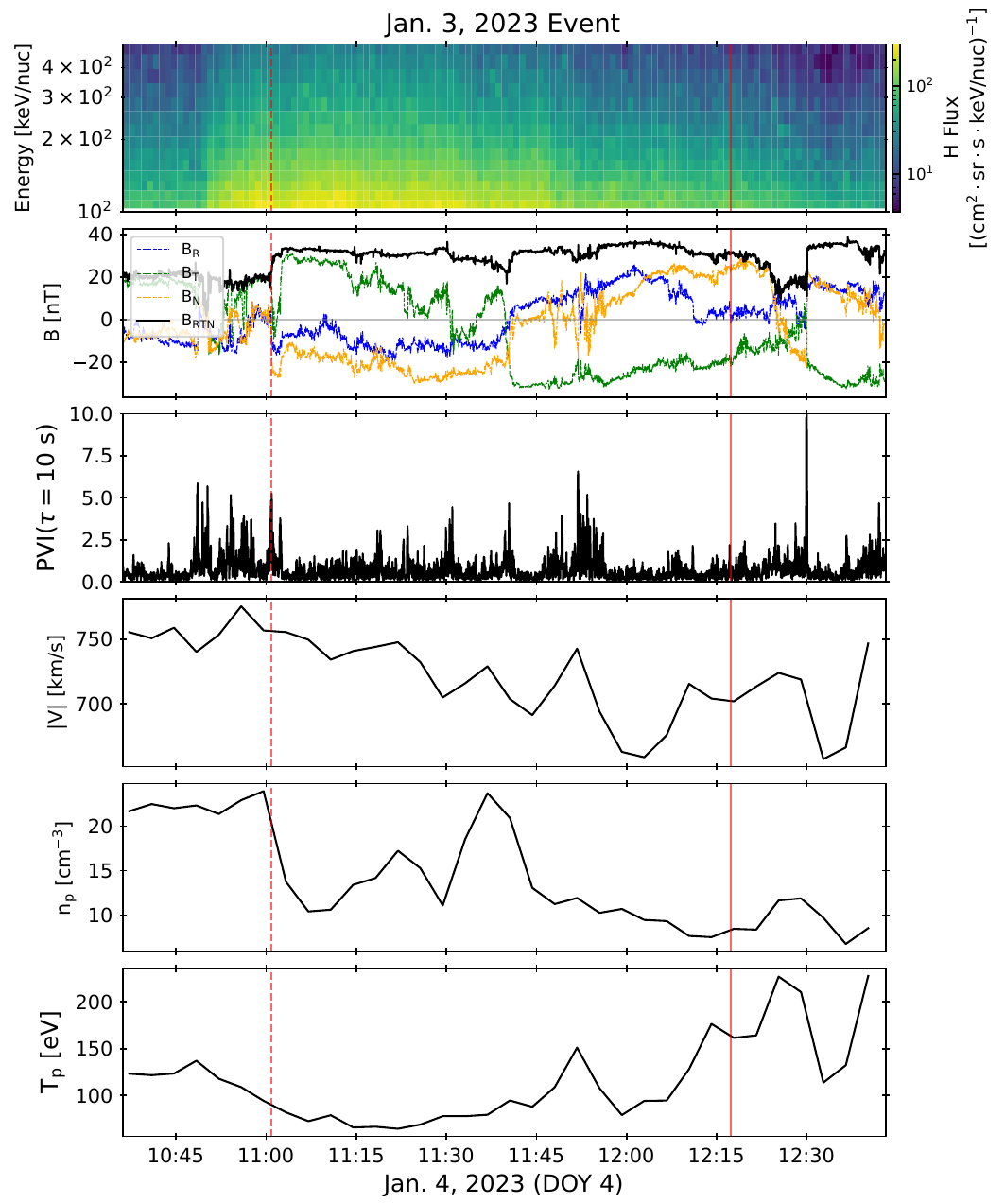}
    \caption{Overview plots for all seven CMEs. Panels are the EPI-Lo Channel P omni-directional proton flux spectrogram, magnetic field, PVI at increment scale \(\tau=10~{\rm seconds}\), bulk velocity, proton density, and proton temperature in descending order for each event. The vertical red dashed line in all panels represents the estimated CME shock arrival time for that corresponding event.}
    \label{fig:overview}
\end{figure}

\newpage
\bibliographystyle{plainnat}





\end{document}